\documentclass[10.5pt,a4paper,oneside]{extarticle}
\usepackage[T1]{fontenc} 
\usepackage[left=25mm,right=20.9mm,top=4.25cm,bottom=3.75cm,headsep=0.25cm,footskip=0.25cm]{geometry}

\usepackage[english]{babel}
\usepackage[affil-it]{authblk}
\usepackage{etoolbox}
\usepackage{lmodern}
\usepackage{setspace}
\usepackage{amssymb,amsthm,amsmath}
\usepackage{tikz}
\usepackage{listings}
\usepackage{color}
\usepackage{caption}
\usepackage{placeins}
\usepackage{booktabs}
\usepackage{enumerate}
\usepackage{graphicx}
\usepackage{titlesec}
\usepackage{mathptmx}
\usepackage{anyfontsize}
\usepackage{t1enc}
\usepackage[utf8]{inputenc}
\usepackage{fancyhdr}
\usepackage{float,xspace}
\usepackage{scalefnt}
	
\usepackage{soul} 
\usepackage{tabulary}
\usepackage[colorlinks=true,linkcolor=blue]{hyperref} 

\usepackage{textcase}

\usepackage[colorinlistoftodos]{todonotes}

\titlespacing*{\section}
{0pt}{12pt}{0pt}
\titlespacing*{\subsection}
{0pt}{12pt}{0pt}
\titlespacing*{\subsubsection}
{0pt}{12pt}{0pt}
\titleformat{\section}
  {\huge\sffamily\Large\bfseries\color{black}}
  {\thesection}{1em}{}
    \titleformat{\subsection}[block]{\bfseries}{\thesubsection}{.5em}{}
    \titleformat{\subsubsection}[block]{\bfseries}{\thesubsubsection}{.5em}{}

\titleformat{\section}{\fontsize{12}{19}\bfseries}{\thesection}{1em}{}

\usepackage{mathptmx} 

\makeatletter

\patchcmd{\@maketitle}{\LARGE \@title}{\fontsize{14}{19.2}\selectfont\@title}{}{} 
\makeatother
\title
{
	\vspace{-5cm}
	\begin{minipage}{\textwidth}	
	\end{minipage}
 \\[0.5cm] \textbf{Performance Analysis Of Binaural Signal Matching (BSM) in the Time-Frequency Domain}
	\author[ ]{Ami Berger$^{(1)}$, Vladimir Tourbabin$^{(2)}$, Jacob Donley$^{(2)}$, Zamir Ben-Hur$^{(2)}$ and Boaz Rafaely$^{(1)}$}
  	\affil[(1)]{School of Electrical and Computer Engineering , Ben-Gurion University of the Negev, Beer-Sheva 84105, Israel}
  	\affil[(2)]{Reality Labs, Meta, 1 Hacker Way, Menlo Park, CA 94025, USA}

}
\date{}

\begin{document}

\clearpage
\setcounter{page}{1}
\maketitle
\thispagestyle{empty}
\fancypagestyle{empty}
{	
	\fancyhf{} \fancyfoot[R]
	{
	}
}


\subsection*{\fontsize{10.5}{19.2}\uppercase{\textbf{Abstract}}}
{\fontsize{10.5}{60}\selectfont The capture and reproduction of spatial audio is becoming increasingly popular, with the mushrooming of applications in teleconferencing, entertainment and virtual reality. Many binaural reproduction methods have been developed and studied extensively for spherical and other specially designed arrays. However, the recent increased popularity of wearable and mobile arrays requires the development of binaural reproduction methods for these arrays. One such method is binaural signal matching (BSM). However, to date this method has only been investigated with fixed matched filters designed for long audio recordings. With the aim of making the BSM method more adaptive to dynamic environments, this paper analyzes BSM with a parameterized sound-field in the time-frequency domain. The paper presents results of implementing the BSM method on a sound-field that was decomposed into its direct and reverberant components, and compares this implementation with the BSM computed for the entire sound-field, , to compare performance for binaural reproduction of reverberant speech in a simulated environment.}

\noindent{\\ \fontsize{11}{60}\selectfont Keywords: Binaural reproduction, BSM, Dynamic-arrays, Head tracking} 

\fontdimen2\font=4pt


\section{\uppercase{Introduction}}
Binaural reproduction of acoustic scenes that are captured by microphone-arrays is becoming increasingly popular \cite{ref1},\cite{ref2},\cite{ref3}, with many applications in teleconferencing and virtual and augmented reality.
A popular method for binaural reproduction involves convolving high order Ambisonics (HOA) signals with the head related transfer function (HRTF) \cite{ref4}. This method is fairly accurate for sufficiently high spherical harmonics (SH) orders, and the incorporation of head tracking enhances the immersion experience of the listener. However, the main drawback of this method lies in the large number of microphones required when the audio signal is recorded by a spherical microphone array, and in the need for a spherical array geometry, which restrict the practical use of this method \cite{ref12}.\newline
In order to reproduce binaural signals using more flexible array geometries, the beamforming-based binaural reproduction (BFBR) \cite{ref5},\cite{ref6},\cite{ref7} method was proposed. In this method the microphone signals are filtered using a set of beamformers, and the output signals are later filtered using the HRTFs and then summed to reproduce the binaural signals. A theoretical framework to set the design parameters of BFBR, such as the look direction and beam number for spherical and planar arrays, was presented in \cite{ref8}. However, for more general array geometries only limited guidelines were suggested, with no guarantee of accurate binaural signal reproduction, and so a comprehensive design methodology is still unavailable. \newline  
With the aim of overcoming the limitations of current beamforming-based methods, and in order to accurately reproduce binaural signals recorded by arrays of arbitrary geometry, the binaural signal-matching (BSM) \cite{ref9},\cite{ref10},\cite{ref11} method was developed. 
BSM refers to the estimation of the binaural signals directly from the array measurements using optimal filters calculated for each ear separately. Recently, the design of a BSM system was described and studied for a varying number of microphones in a semi-circular array \cite{ref13} incorporating head tracking \cite{ref21},\cite{ref23}; it was shown that the accuracy of BSM is sensitive to the position of the microphones rather than to their number. In particular, it was shown that the closer a microphone in the array is to an ear, the better the binaural reproduction in that ear. Therefore, the main drawback of the BSM method is that it performs poorly for high frequencies, especially in cases where one of the ears is relatively far from all the microphones of the array. In order to improve the perceptual performance it was proposed to use Magnitude Least-Squares (MagLS) \cite{ref29} instead of Least-Squares (LS) for high frequencies. However, while the use of MagLS improved the overall perceptual experience of a reproduced binaural signal, BSM still performs perceptually badly when one of the ears is relatively far from all of the array microphones.\newline 
Parametric spatial audio and binaural reproduction have also been studied as an alternative to the BSM and beamforming methods described above. In this approach the sound-field is decomposed into components, typically direct sources and reverberant parts, and each is estimated and reproduced separately \cite{ref22},\cite{ref24},\cite{ref25},\cite{ref26},\cite{ref27}. While these approaches show promising performance, the latter may depend on estimation accuracy, which becomes challenging in complex environments. Nevertheless, the parametric approach also motivated the study in this paper. 
\newline
With the limitations of previous method in mind, this paper aims to study and analyze the performance of BSM with a parameterized sound-field in the time-frequency domain. Specifically, an acoustic scene that is separated into direct and reverberant components is investigated. The paper investigates the potential of improving BSM by incorporating sound-field parameterization, especially in cases where BSM currently fails.


\section{\uppercase{Mathematical Background}}
\label{Math-back}
This section provides the mathematical model of BSM. Throughout the paper, the standard spherical coordinate system is used, denoted by $(r,\theta,\phi)$, where $r$ is the distance to the origin, $\theta$ is the elevation angle measured from the Cartesian z axis downwards to the Cartesian xy plane, and $\phi$ is the azimuth angle measured from the positive x axis towards the positive y axis.
A sound-field can be described as a linear combination of plane-waves $a(k,\theta,\phi)$, where  $k=\frac{2\pi}{\lambda}$ is the wave-number, $\lambda$ is the wave-length and $(\theta,\phi)$ is the wave arrival direction.
Further, assume that the sound-field is composed of $L$ far-field sound sources arriving from directions ${(\theta_l,\phi_l)}^L_{l=1}$, with source signals ${s_l(k)}^L_{l=1}$ \cite{ref15}. The sound-field is captured by an array of $M$ microphones, which are located at ${(r_m,\theta_m,\phi_m)}^M_{m=1}$, centered at the origin. The noisy array measurements can be described by the following narrow-band model \cite{ref15}:
\begin{equation}
    \mathbf{x}(k)=\mathbf{V}(k)\mathbf{s}(k)+\mathbf{n}(k)
    \label{eq:1}
\end{equation}
where $\mathbf{x}(k)=[x_1(k),x_2(k),...,x_M(k)]^T$ is the microphone-signal vector (measurements),\newline $\mathbf{V}(k)=[\mathbf{v}(k,\theta_1,\phi_1),\mathbf{v}(k,\theta_2,\phi_2),...,\mathbf{v}(k,\theta_L,\phi_L)]$ is an $M\times L$ complex matrix with columns $\mathbf{v}(k,\theta_l,\phi_l)$ representing the array steering vector from the $l$-th source to the microphone positions for all $l=1,2,...,L$ sources, \newline$\mathbf{s}(k)=[s_1(k),s_2(k),...,s_L(k)]^T$ is the source-signal vector, and $\mathbf{n}(k)$ is an additive-noise vector.
In the BSM model we further assume that a listener is positioned with the center of the head coinciding with the origin,  where $h^{l,r}(k,\theta,\phi)$ denotes the HRTF of the left and right ears of the listener using the superscripts $(\cdot)^l$ for the left ear and $(\cdot)^r$ for the right ear.
The signal at the left and rights ears can now be written as \cite{ref13}:
\begin{equation}
    p^{l,r}(k)=[\mathbf{h}^{l,r}(k)]^T\mathbf{s}(k)
\end{equation}
where $\mathbf{h}^{l,r}=[h^{l,r}(\theta_1,\phi_1),h^{l,r}(\theta_2,\phi_2),...,h^{l,r}(\theta_L,\phi_L)]^T$ contains the HRTFs corresponding to the directions of the sources.\newline
Next, assume that the configuration of the microphone array is known, such that the steering matrix $\mathbf{V}(k)$ can be calculated analytically or numerically. In the first step, the array measurements are filtered and combined, in a similar manner to beamforming:
\begin{equation}
    z^{l,r}(k)=[\mathbf{c}^{l,r}(k)]^H\mathbf{x}(k)
    \label{eqn:3}
\end{equation}
where $\mathbf{c}$ is an $M\times1$ complex vector holding the filter coefficients. Next, $\mathbf{c}$ is chosen to minimize the following mean-squared error between $z^{l,r}(k)$ and $p^{l,r}(k)$, the binaural signals in (2), for each ear separately:
\begin{equation}
    err^{l,r}_{bin}(k)=\mathbb{E}[|p^{l,r}(k)-z^{l,r}(k)|^2]
\end{equation}
where $\mathbb{E}[\cdot]$ is the expectation operator.
Next, assume that the noise is uncorrelated to the sources. This leads to the following error formulation:
\begin{equation}
\label{eqn:5}
    err^{l,r}_{bin}(k)=\|[\mathbf{s}(k)]^H([\mathbf{V}(k)]^H\mathbf{c}^{l,r}(k)-[\mathbf{h}^{l,r}(k)]^*)\|^2_2+\|[\mathbf{n}(k)]^H\mathbf{c}^{l,r}(k)\|^2_2
\end{equation}
Minimizing the error in (\ref{eqn:5}) leads to the following solution:
\begin{equation}
    \mathbf{c}^{l,r}_{opt}(k)=(\mathbf{V}\mathbf{R_s}\mathbf{V}^H+\mathbf{R_n})^{-1}\mathbf{V}\mathbf{R_s}[\mathbf{h}^{l,r}]^*
    \label{eqn:6}
\end{equation}
where $\mathbf{R_s}=E[\mathbf{s}\mathbf{s}^H]$, $\mathbf{R_n}=E[\mathbf{n}\mathbf{n}^H]$.
Next, assume that the sources are uncorrelated as in \cite{ref21}, leading to \newline$\mathbf{R_s}=\sigma_s^2\mathbf{I}_L$, and also that the noise in uncorrelated between microphones, such that $\mathbf{R_n}=\sigma_n^2\mathbf{I}_M$. These assumptions lead to the following simplification of (\ref{eqn:6}):
\begin{equation}
    \mathbf{c}^{l,r}_{opt}(k)=(\mathbf{V}\mathbf{V}^H+\frac{1}{SNR}\mathbf{I}_M)^{-1}\mathbf{V}[\mathbf{h}^{l,r}]^*
    \label{eqn:7}
\end{equation}
where $SNR=\frac{\sigma_s^2}{\sigma_n^2}$. Equation (\ref{eqn:7}) has been used in \cite{ref21}. The advantage of (\ref{eqn:7}) is that it does not require the estimation of $\mathbf{R_s}$, but this in turn may lead to reduced performance. The next section presents an approach to enhance performance by incorporating limited information on $\mathbf{R_s}$.


\section{\uppercase{BSM with parameterized sound-field}}
\label{section:BSM-param}
This section investigates the BSM method described in \cite{ref21}, but extended to incorporate a parameterization of the sound-field. The parameterization is based on the assumption that the measured sound-field can be decomposed into two components as follows:  
\begin{equation}
    \mathbf{x}(n,k)=\mathbf{x}_{d}(n,k)+\mathbf{x}_{r}(n,k)+\mathbf{n}(n,k)
    \label{eqn:8}
\end{equation}
where $\mathbf{n}(k)$ is an additive-noise vector, and $\mathbf{x}_{d}(n,k)$ represents the direct signal from the sound source in the time-frequency domain, modeled as a single far-field plane wave, written as:
\begin{equation}
    \mathbf{x}_{d}(n,k)=\mathbf{v}(k,\theta_d,\phi_d){s}_{d}(n,k)
\end{equation}
where $(\theta_d,\phi_d)$ represents the DOA of the direct signal, and ${s}_{d}(n,k)$ represents the source signal.
$\mathbf{x}_{r}(n,k)$ represents the reverberant part of the measured signal in the time-frequency domain and is typically composed of a large, unknown number of sources that arrive from unknown directions. This model can represent a single source in a room where $\mathbf{x}_{d}$ are the measurements of the direct sound from the source and $\mathbf{x}_{r}$ are the measurements of the source reflections from room boundaries.
There are methods available that estimate $\mathbf{x}_{d}(n,k)$ and its direction-of-arrival (DOA) for each time-frequency bin \cite{ref22},\cite{ref27},\cite{ref28}, but in this paper $\mathbf{x}_{d}(n,k)$ and its DOA are assumed to be known. Knowing $\mathbf{x}_{d}(n,k)$, there are methods that can reproduce the direct component of the binaural signal, $p^{l,r}_d(n,k)$ \cite{ref21},\cite{ref22},\cite{ref8},\cite{ref1}. Nevertheless, in this paper the BSM method was chosen  as this method is the focus of this paper. Reproducing the binaural signal of the reverberant component, $p^{l,r}_r(n,k)$, from the reverberant component of the measurement, $\mathbf{x}_{r}(k)$, is a more challenging problem since it usually requires an estimation of $\mathbf{R_s}$. The simplification of $\mathbf{R_s}$ applied in (\ref{eqn:7}) assumes that the sources are uncorrelated, e.g. a diffuse sound-field, and so it could be a reasonable approximation for the reverberant part.\newline 
In summary, computing the BSM filters separately for the direct and reverberant components is expected to produce better results compared to computing the BSM filters for the entire sound-field, because in the former the direct sound components are expected to be reproduced more accurately. These components are formulated as:
\begin{equation}
    \hat{p}^{l,r}(n,k)=\hat{p}^{l,r}_d(n,k)+\hat{p}^{l,r}_r(n,k)
    \label{eqn:9}
\end{equation}
where $\hat{p}^{l,r}_d(n,k)$ and $\hat{p}^{l,r}_r(n,k)$ are the reproduced binaural signals of the direct and reverberant components using the BSM method, as described in (\ref{eqn:7}). The performance of BSM with the proposed decomposition is evaluated in this paper numerically using binaural signal error and perceptually, using a listening test.


\section{\uppercase{Simulation study}}
This section presents a simulation-based analysis of the performance of the proposed BSM method with sound-field decomposition, compared to the BSM method computed from the entire measured signal without applying decomposition. 
\subsection{\uppercase{set-up}}
\label{section:set-up}
\begin{figure}[H]
\centering
\includegraphics[width=4cm]{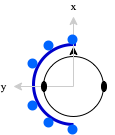}
\caption{An illustration of a head and the semi-circular array where the blue dots represent the array microphones, and the $x$ and $y$ grey arrows represent the positive $x$ and $y$ axes, respectively.}
\label{fig:illustration1}
\end{figure}

A point source was simulated inside a room of dimensions $8\times 5\times 3\,$m, having a reverberation time of ${T_{60}=0.68\,}$s, using the image method \cite{ref19}. The source location in the room was $(2.47,2.27,1.7)\,$m. The source signal was a 5s long recording of female speech, taken from the TIMIT database \cite{ref20}, sampled at $48\,$kHz. A semi-circular microphone array was centered at $(2,2,1.7)\,$m, with a DRR value of $4.5\,$dB, compromising $M=6$ omni-directional microphones arranged on the horizontal plane. Microphone positions were denoted using spherical coordinates by $(r_m,\theta_m,\phi_m)$ for $m=1,...,M$ relative to the array center, where $r_m=10\,$cm, $\theta_m=\frac{\pi}{2}\,$rad and $\phi_m=\pi-\frac{\pi(m-1)}{M-1}\,$rad. Microphone measurement signals are denoted by $\mathbf{x}(t)=[x_1(t),x_2(t),...,x_M(t)]^T$. The component of $\mathbf{x}(t)$ representing the direct contribution from the source, $\mathbf{x}_d(t)$ as in (\ref{eqn:8}), was calculated by assuming a free-field environment.   
The HRTF in the simulation was taken from the Cologne database \cite{ref18} measurements of the Neumann KU100 manikin, sampled at a frequency of $48\,$kHz. The HRTFs for the DOAs of the assumed sources were interpolated in the SH domain using a SH order of 30. 
The head was centered at $(2,2,1.7)\,$m and was aligned with the positive $x$ axis. An illustration of the array position relative to the head position and alignment is presented in Figure \ref{fig:illustration1}. Assuming the semi-circular array represents an array on AR glasses, for example, the selected orientation of the array relative to the head was chosen as this was found to be the most challenging for the BSM algorithm \cite{ref21}.  

\subsection{\uppercase{methodology}}
 Having computed $\mathbf{x}(t)$ and $\mathbf{x}_d(t)$ as described above, $\mathbf{x}(n,k)$ and $\mathbf{x}_{d}(n,k)$ were computed using the Short-Time-Fourier-Transform (STFT), with a Hamming window of $32\,$ms and a hop length of $16\,$ms. $\mathbf{x}_{r}(n,k)$ was then calculated using (\ref{eqn:8}).
Using (\ref{eqn:7}), two BSM filters were computed, one for the direct component and one for the reverberant component of the signal, denoted $\mathbf{c}^{l,r}_{d}(k)$ and $\mathbf{c}^{l,r}_{r}(k)$, respectively. As described in section \ref{Math-back}, the BSM method was developed assuming a sound-field composed of $L$ far-field sources. When computing the filter for the reverberant component, it was assumed, similarly, that the sound-field is composed of $L=240$ sources with DOAs that correspond to a spirally-nearly-uniform distribution \cite{ref16}. A single source with $L=1$ and a DOA of $(\theta_d=\frac{\pi}{2},\phi_d=\frac{\pi}{6})$, relative to the array's center, was assumed when computing the filter of the direct component.
The steering vectors for the reverberant and direct components were generated analytically in the SH domain according to section 4.2 in \cite{ref17} using the corresponding number of $L$ sources, the array geometry and the DOAs of the assumed sources. Next, $SNR=20\,$dB was assumed for the reverberant components, while $SNR=\infty$ was assumed for the direct component, and both values were substituted in (\ref{eqn:7}). The BSM weights of the reverberant component were calculated using MagLS \cite{ref29} for frequencies in the range of $[1500,24000]\,$Hz.\newline
Next, $\hat{p}^{l,r}_r(n,k)$ and $\hat{p}^{l,r}_d(n,k)$, representing the output of the BSM method for the reverberant and direct components, were calculated according to (\ref{eqn:3}) using $\mathbf{c}^{l,r}_{r}(k)$,$\mathbf{x}_{r}(n,k)$ and $\mathbf{c}^{l,r}_{d}(k)$,$\mathbf{x}_{d}(n,k)$, respectively.\newline
$\hat{p}^{l,r}_{BSM}(n,k)$, representing the solution of implementing the BSM method directly from the array measurements without sound-field decomposition, were calculated according to (\ref{eqn:3}) using $\mathbf{c}^{l,r}_{r}(k)$ and $\mathbf{x}(n,k)$.\newline
${p}^{l,r}_{ref}(n,k)$, the reference signals at the ears, were calculated by convolving the HRTFs of the left and right ears with the HOA signals of order 14 that were calculated using the image method as described in section \ref{section:set-up}.\newline
${p}^{l,r}_{d-ref}(n,k)$, the direct component of the reference signals at the ears, were calculated by convolving the HRTFs of the left and right ears with the HOA signals of order 14 that were calculated by assuming a free-field environment as described in section \ref{section:set-up}.\newline
${p}^{l,r}_{r-ref}(n,k)$, the reverberant component of the reference signals at the ears, were calculated using (\ref{eqn:9}).\newline
The normalized mean-squared error (NMSE) of the binaural signals was calculated as:
\begin{equation}
    NMSE(k)=\frac{\mathbb{E}_n[|\hat{p}^{l,r}(n,k)-{p}^{l,r}_{ref}(n,k)|^2]}{\mathbb{E}_n[|{p}^{l,r}_{ref}(n,k)|^2]}
\end{equation}
where ${p}^{l,r}_{ref}(n,k)$ is the reference binaural signal of a certain time-frequency bin, and $\hat{p}^{l,r}(n,k)$ is the reproduced binaural signal of a certain time-frequency bin.

\subsection{\uppercase{Simulation results}}
\begin{figure}[H]
\centering
\includegraphics[width=14cm]{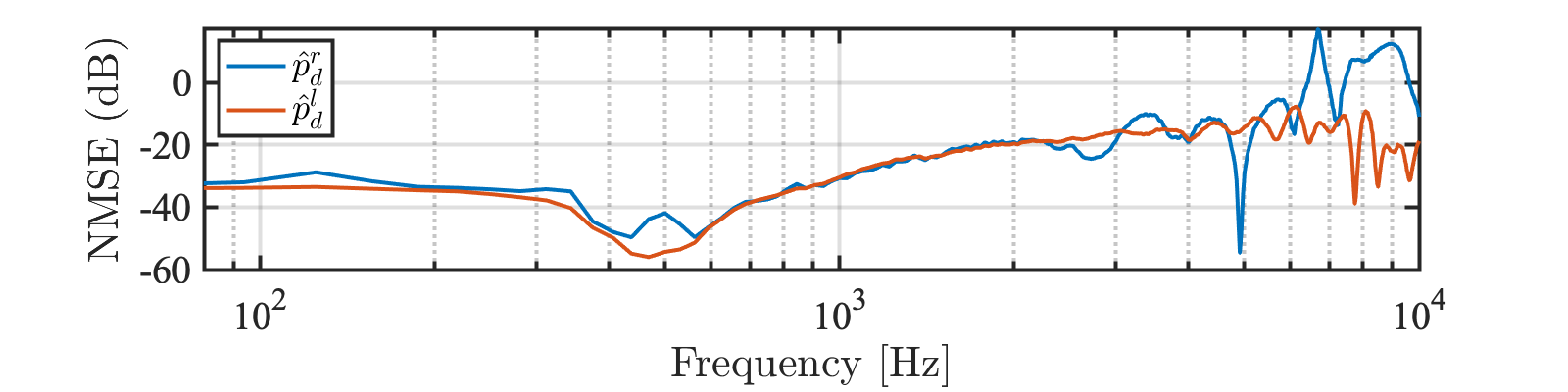}
\caption{The NMSE of the direct component of the binaural signal for the left and right ears ($p_d^l, p_d^r$), calculated using ${p}^{l,r}_{d-ref}$ as the reference binaural signals.}
\label{fig:figure2}
\end{figure}

\begin{figure}[H]
\centering
\includegraphics[width=14cm]{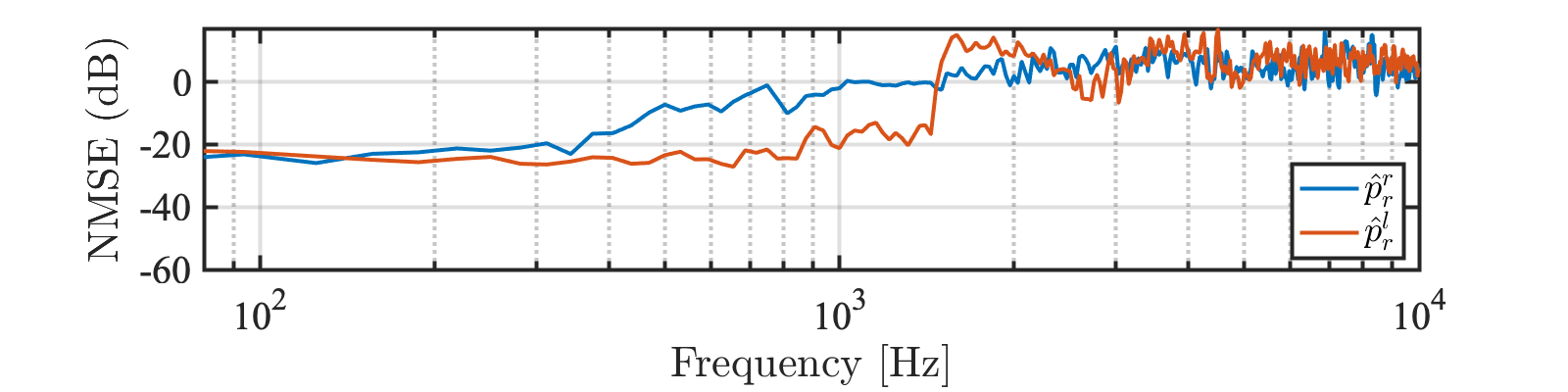}
\caption{The NMSE of the reverberant component of the binaural signal for the left and right ears ($p_r^l, p_r^r$) calculated using ${p}^{l,r}_{r-ref}$ as the reference binaural signals.}
\label{fig:figure3}
\end{figure}

\begin{figure}[H]
\centering
\includegraphics[width=14cm]{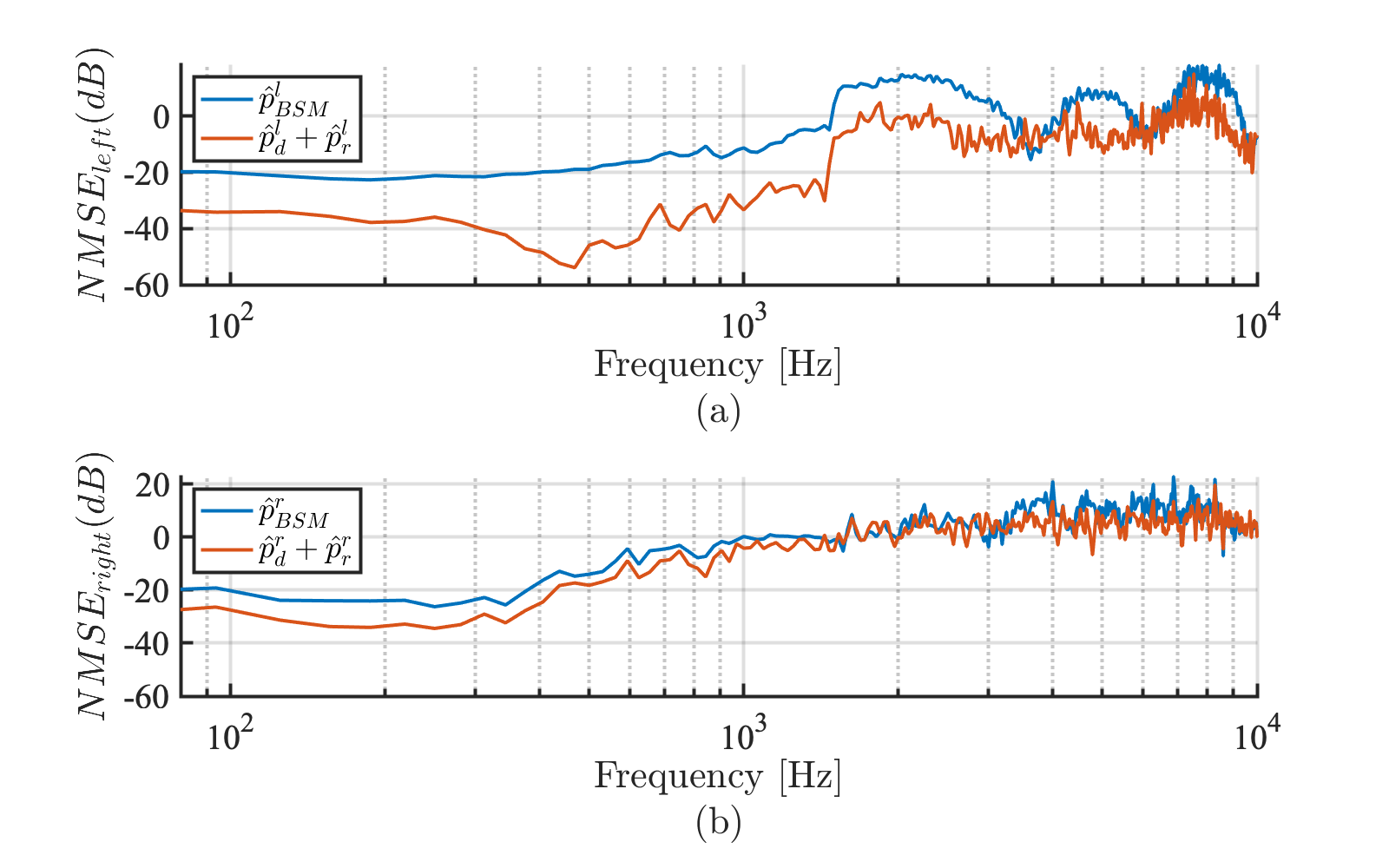}
\caption{The NMSE of the left (a), and right (b) ears calculated using $\hat{p}^{l,r}_{ref}$ as the reference binaural signals.}
\label{fig:figure1}
\end{figure}

In order to study the performance of the BSM method when applied to a decomposed sound-field, the NMSE of the direct and reverberant components were calculated and are presented in Figures \ref{fig:figure2} and \ref{fig:figure3}. Figure \ref{fig:figure2} shows that the NMSE of the reproduced direct component of the binaural signal is relatively low, for both the left and the right ears, in particular at the low frequencies. This result indicates a fairly accurate reproduction of the direct component, which agrees with the expected performance of the BSM method designed for a sound-field compromised of $L<M$ sources as shown in \cite{ref13}.\newline
The NMSE of the reverberant component is presented in Figure \ref{fig:figure3}. As can be observed, the NMSE is higher at higher frequencies and lower for the ear closest to the microphone positions (in this simulation the left ear), see Figure \ref{fig:illustration1}. These results agree with the expected performance of the BSM method as described in \cite{ref21}.\newline
The NMSE of the reproduced binaural signal using the BSM method without sound-field decomposition, $\hat{p}^{l,r}_{BSM}(n,k)$, and the reproduced binaural signal using the BSM method with sound-field decomposition, $\hat{p}^{l,r}_{r}+\hat{p}^{l,r}_{d}$, were also calculated and are presented in Figure \ref{fig:figure1}. As can be observed there is a clear overall improvement in the NMSE of the reproduced binaural signals for the BSM method with sound-field decomposition, compared to the standard BSM method, especially for the left ear (which is closer to the array). This demonstrates that a more accurate reproduction of the binaural signal using the BSM method can be achieved by sound-field decomposition.\newline
By comparing the NMSE of the direct and reverberant components, we can deduce that the NMSE is dominated by the reverberant component. This simulation study shows the potential for sound field-decomposition - if such decomposition can be implemented in practice, it can significantly improve the performance of the BSM algorithm. 

\section{\uppercase{listening experiment}}
\begin{figure}[H]
\centering
\includegraphics[width=10cm]{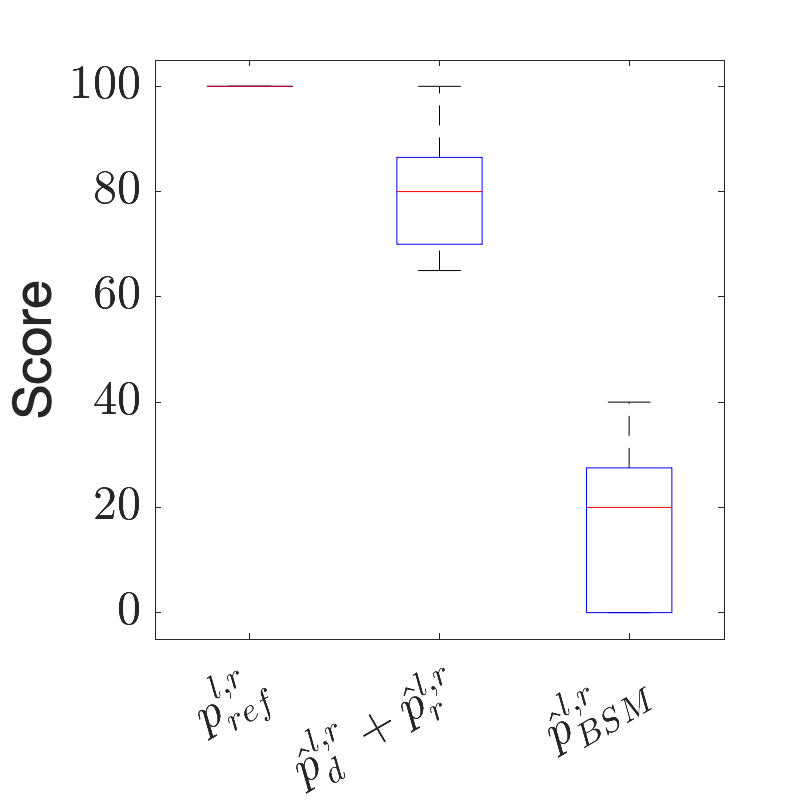}
\caption{Box-plot of the listening experiment score. The median is marked by a red line, the 25th and 75th percentiles are marked as the bottom and top blue edges respectively, and the maximal and minimal grades are marked by black lines. The scores are based on overall quality.}
\label{fig:figure4}
\end{figure}

A preliminary listening test was reproduced with the aim of evaluating the similarity between the reference binaural signal ${p}^{l,r}_{ref}$, and the reproduce signals $\hat{p}^{l,r}_{BSM}$ and $\hat{p}^{l,r}_{r}+\hat{p}^{l,r}_{d}$. The experiment was based on the MUltiple Stimuli with Hidden Reference and Anchor (MUSHRA) test \cite{ref30} and incorporated 9 normal hearing participants. The preliminary listening test scores are presented using a box-plot in Figure \ref{fig:figure4}. As can be observed in Figure \ref{fig:figure4}, the scores of the reproduced binaural signal for the BSM with sound-field decomposition are higher than the scores of the standard BSM method. These results agree with the results presented in Figure \ref{fig:figure1}; in particular, the score of the BSM with sound-field decomposition was very high and close to the reference, suggesting that this approach could also provide improvement that is perceptually notable.    

\section{\uppercase{Conclusions}}

In this paper, binaural reproduction with the BSM method was studied with a semi-circular array and a decomposed sound-field. It was shown that the perceptual and overall accuracy of the reproduced binaural signal using a decomposed sound-field is higher than in the case of using the standard BSM method, especially where the distance between the listener ear position and the microphones is large. A listening test demonstrated that the accurate reproduction of the direct component was indeed important for perception. Future work could include the incorporation of a spatial coding method to implement sound-field decomposition, the development of a design framework that better reproduces the reverberant component, and an extension of the listening test performed in this work. Future work could also include the study of the BSM method with other array configurations, and the development of a design framework that improves binaural reproduction.



\bibliographystyle{abbrv}
\renewcommand{\refname}{\normalfont\selectfont\normalsize}

\end{document}